# Socio-Technical Root Cause Analysis of Cyber-enabled Theft of the U.S. Intellectual Property - The Case of APT41

Mazaher Kianpour
mazaher.kianpour@ntnu.no
Norwegian University of Science and Technology
Gjøvik, Norway

## ABSTRACT

Increased connectivity has made us all more vulnerable. Cyberspace, besides all its benefits, spawned more devices to hack and more opportunities to commit cybercrime. Criminals have found it lucrative to target both individuals and businesses, by holding or stealing their assets via different types of cyber attacks. The cyber-enabled theft of Intellectual Property (IP), as one of the most important and critical intangible assets of nations, organizations and individuals, by foreign countries has been a devastating challenge of the United States (U.S.) in the past decades. In this study, we conduct a socio-technical root cause analysis to investigate one of the recent cases of IP theft by employing a holistic approach. It concludes with a list of root causes and some corrective actions to stop the impact and prevent the recurrence of the problem in the future. Building upon the findings of this study, the U.S. requires a detailed revision of IP strategies bringing the whole socio-technical regulatory system into focus and strengthen IP rights protection considering China's indigenous innovation policies. It is critical that businesses and other organizations take steps to reduce their exposure to cyber attacks. It is particularly important to train employees on how to spot potential threats, and to institute policies that encourage workers to report potential security failures so that action can be taken quickly. Finally, we discuss how cyber ranges can provide an efficient and safe platform for dealing with such challenges. The results of this study can be expanded to other countries in order to protect their IP rights and deter or prevent and respond to future incidents.

## 1 INTRODUCTION

After an "amazing and productive meeting", so-called by Donald Trump, between U.S. President and China's Leader, the White House announced that it is expected that the two nations immediately begin negotiations on structural changes with respect to forced technology transfer, intellectual property protection, cyber intrusions, and cyber theft [13]. Despite Trump's statement, the outcome of the negotiations was highly uncertain as in September 2019, the U.S. Deputy Assistant Attorney General stated that more cases are being opened that implicate trade secrets and intellectual property theft, and most of them are attributed to China. Moreover, the Department of Justice warn companies to bolster their defenses [14], and the director of Federal Bureau of Investigation (FBI) said that FBI has nearly 1,000 investigations open into economic espionage and attempted intellectual property theft, nearly all of them leading back to China [27].

According to a survey by CNBC, 1 in 5 American corporations say China has stolen their intellectual properties within the last year [29]. Intellectual properties - patent, trademarks, trade secrets, are the pillars of a creative and innovative economy and have a vital role in the development of countries all over the world. According to Global Innovation Policy Center, improving IP protection leads to the following: [5]:

- 55% more likely to adapt to sophisticated, state-of-the-art technology
- 53% more likely to employ high-skilled and high-paid workers
- 53% more likely to experience increased R&D activity
- 30% more likely to attract venture capital and private equity funds
- 39% more likely to attract foreign investment
- Over 4 times more online and mobile content generated

**Table 1: Steps of root cause analysis in this study (Adopted from [2])**

| Step | Purpose | Outputs | Tools |
|------|---------|---------|-------|
| Define the Problem | Understanding and scoping the problem and describing the existing environment in where the problem occurs, and finding the impacts of the problem | Problem Definition | Survey |
| Find Causes | Better understanding of the problem and creating a broad overview of possible causes | List of possible causes | Socio-technical Systems Theory and Fishbone Diagram |
| Find the root causes | Identifying the root causes leading to the incident | Description of the root causes | Security by Consensus Model |
| Find Solutions | Propose feasible solutions to eliminate the root causes and assess the potential effectiveness of the solution | Description of the Solutions and a list of prioritized solutions | Cybersecurity Value Chain and Q Methodology |

On the other hand, calculation of the cost of cyber-attacks is affected by numerous factors and it is virtually impossible to determine the precise costs that individuals, organizations and governments may incur. IP theft cyber-crimes are no exception. The estimation of costs varies regarding the size of the economic loss attributed to the theft of IP and trade secrets. Estimated annual cost to the U.S. economy through the theft of trade secrets exceeds $225 billion and could be as high as $600 billion [6]. These numbers show that the reported damage from the NotPetya cyber-attack, which is estimated at $10 billion and known as the costliest cyber-attack to date [8], is much smaller in comparison to the IP thefts against a single country in one year.

The IP protection benefits along with the high cost of IP theft necessitate a societal response. For a societal response to be effective, understanding the problem and finding its root causes need to be one of the first steps. This study presents the work undertaken to investigate the socio-technical causes of failure in Intellectual Property (IP) protection in the U.S. This research could be a foundation for more substantial and larger-scale research and studies to develop a significant base to examine the existing practices and the underlying causes of failure across a range of individual, organizational and governmental in other countries. We acknowledge the multiplicity of the interdependent causes of failure in IP protection and take this into account through the use of multiple disciplines and field of study such as politics, law and economics.

The rest of the paper is structured as follows. Section 2 describes our research method and the steps we took to analyze the root causes of this problem. The analysis results, the employed tools and the proposed corrective actions are discussed in Section 3. Finally, Section 4 concludes the paper.

## 2 RESEARCH METHOD

IP theft exists within a broad range of scope, persistence, and severity across different industries and sectors. Some of these incidents cause a minor nuisance, others cause financial and some can damage reputations of whole industries and even nations. In most of these cases, prevention efforts are preferable to dealing with the consequences of them. Hence, root cause analysis of these problem can be the key to setting up a prevention strategy to IP theft . In terms of scope and extent of a root cause analysis, there can be a large variation depending on the nature of problem and its repercussions. Root cause analysis is part of an complete corrective process, and identifying the root causes is only one part of this process.

To our knowledge, we could not find a commonly accepted definition and approach of root cause analysis. This analysis is understood as a wide range of approaches, tools and techniques used to identify causes of problems and eliminate them to prevent from reoccurring those problems [2, 10]. However, we defined the root causes *as the casual or contributing factors that, if corrected would, with a given degree of certainty, prevent recurrence of the identified problem*. Individuals, organizations and governments would save time and money by eliminating the root cause. It also prevents the problems in other areas and improves the communication between groups and actors. Basically, the root cause analysis should be performed all the time when problems occur. Problems that are left unattended tend to become crisis. Some of the reasons why problems occur include supplier defects, under/overspending budgets, human error, out of control processes, etc. Consequently, to conduct a root cause analysis, we took the steps outlined in Table 1. Although this steps are presented as a sequential process, in reality we repeated some steps to revise and fine-tune the study.

## 3 SOCIO-TECHNICAL ROOT CAUSE ANALYSIS OF INTELLECTUAL PROPERTY THEFT

In this section, we follow the aforementioned steps to conduct a socio-technical root cause analysis of IP theft by Chinese state-sponsored groups against the U.S. cancer research centers. These operations are known as Advanced Persistent Threat (APT) and this case is categorized under APT41.

### Define the problem

Economic espionage is not a new problem but with the development of the Internet, new means and methods are being developed at an accelerating rate. Espionage is a national security crime violating Title 18 USC, 792-798[1] and Article 106a, Uniform Code of Military Justice (UCMJ)[2]. Convictions of espionage require the transmitting, collecting and losing national defense information with intent to aid a foreign power or harm the U.S. Although in many resources economics espionage and IP theft are used interchangeably, they are different. Economic espionage and IP theft are defined under 1831 and 1832 of the Economic Espionage Act of 1996[3], respectively. Economic Espionage comprises behavior that denies the rightful owner of the economic benefit of property that the owner has gone to reasonable means to protect and does so with the intent to benefit a foreign entity. Whereas, IP theft covers the conversion of a an intellectual property to the economic benefit of anyone other than the rightful owner. In fact, all of these are relevant and governments and organizations should consider all of them. However, different violations may result in different courses of action against the perpetrators. It also should be noted that, accurately determination of the end users is not always possible. What appears to be a domestic perpetrator may in fact turn out to be a foreign collection effort.

With the advent of information age and cyberspace, cyber-enabled methods allowed adversaries to harm their targets far disproportionate to their own size or resources. These malicious cyber-enabled actions are called differently, depending on their purpose in various contexts, such as cyber-crime, cyberespionage, cybersabotage and cyberterrorism. Reports show that the U.S. government, businesses, industries and banks have been the target of numerous small- and large-scale cyber attacks which is described as the greatest transfer of wealth in the history of this country [1, 31]. Hence, the Foundation for Defense of Democracies (FDD) introduced a new class of these security threats labeled as Cyber-Enabled Economic Warfare [28]. These threats arise from opportunities provided by rapidly evolving world's features. The ever-changing nature of communications, and growing technology-driven, information-intensive societies make the tackling these threats a must.

Today, digital technologies and Internet file sharing networks has facilitated intellectual property theft. The U.S. believes that China is responsible for over 90 percent of cyber-enabled intellectual property theft in the United States [12]. In Table 2, we summarize the significant IP thefts by Chinese against the U.S. businesses and organizations in 2019. The incident reported in August is related to the theft of protected data from multiple U.S. cancer institutes by Chinese state-sponsored hackers as a part of APT41 operations. Healthcare research centers have been frequently targeted by Chinese cyber attacks in last decades. Cancer-related research institutes, in particular, are becoming more popular targets for Chinese APT groups reflecting the growing concerns of China over dramatically increasing cancer incidence and mortality rates, and Five-Year economic development plans[4].

APT41, APT22, APT18 and APT10 are the most significant identified Chinese cyber espionage operations against the U.S. sectors including telecommunication, healthcare, and high-tech. In this study, we focused on APT41 which through 2014 started to carry out an array of IP theft operations against 11 industries, particularly healthcare and pharmaceutical, in 14 countries. The activity of this group traces back to 2012 by conducting financially motivated operations against the video gaming industry. However, since 2014, they expanded into a state-sponsored group targeting more industries and focusing on espionage activities. The group's noticeable use of supply chain compromises, tracking individuals and conducting surveillance highlights a creative and well-resourced adversary. A comprehensive report by FireEye demonstrates that APT41 group also targeted third parties and leveraged this access to target additional victims. This report also outlines the technical information and the links to other known Chinese cyber espionage operators. According to this report, APT41 leveraged timely news stories as content to lure in their spear-phishing emails, although social engineering content does not always correlate with targeted users or organizations. Moreover, access to third-parties and compromising supply chain enabled APT41 to develop tactics, techniques, and procedure (TTPs) that were later leveraged to target additional victims. A comprehensive

---

**Table 2: The Significant IP Thefts By China against the U.S. in 2019 [19]**

| Month | Descriptionn |
|-------|-------------|
| October | Chinese hackers engaged in a multi-layer campaign between 2010 and 2015 to acquire IP from foreign companies to support the development of the Chinese C919 airliners. |
| August | Chinese state-sponsored hackers were revealed to have targeted multiple U.S. cancer institutes to take information relating to cutting edge cancer research. |
| July | State-sponsored Chinese hackers conducted a spear-phishing campaign against employees of three major U.S. utility companies |
| June | Over the course of seven years, a Chinese espionage group hacked into ten international cellphone providers operating across thirty countries to track dissidents, officials, and suspected spies. |
| May | Hackers affiliated with the Chinese intelligence service reportedly had been using NSA hacking tools since 2016, more than a year before those tools were publicly leaked. |
| January | U.S. prosecutors unsealed two indictments against Huawei and its CFO Meng Wanzhou alleging crimes ranging from wire and bank fraud to obstruction of justice and conspiracy to steal trade secrets. |

list of enterprise techniques used by APT41 are collected by MITRE ATT&CK[5].

## Find causes

A well-executed and effective IP strategy needs addressed policies at all levels of socio-technical systems. This includes various communities, all types of national and international institutions, social organizations and governments. It can be argued that even sectors that are not typically associated with an IP strategy in an institution can complement an IP development-oriented policy, impacting other communities and institutions. A wide range of actors either provide or interact with IP protection. We identified these different actors at each level of a socio-technical system (i.e. international, national, organizational and individual) and studied their role in this case both in the U.S. and China. It should be noted that actors are not constrained within each level and can operate with multiple actors in other levels.

[5]https://attack.mitre.org/groups/G0096/

To go even further, we used a four level construct as an instrument to study the casual factors involved in IP theft. As such, these levels can be used as a generalized tool for analysis: policy level where the statement of intent and long-term objectives are described and implemented as a procedure or protocol; a strategic level where the actors are set up with determined actions to achieve the goals; an operational level where connects the details of tactics with the goals of strategy and coordinates the different tasks within an individual actor, and a tactical level where the specific techniques and procedures are organized, employed and executed for each task. This delineation will be used to provide casual factors of IP theft.

**Policy Level**. There are two global perspectives on IP rights protection: one group considers IP rights protection as an obstacle to domestic developments by creating barriers to the use of intangible resources on favorable terms [7], and the other group considers them as a means to foster growth in domestic industries, encourage innovation and protect foreign firms in high-infringement jurisdictions [21]. These two different views on whether and how IP rights promote development in domestic and global economies often result in policies that are either conductive to development or are challenging as development aids. Nevertheless, we will continue with describing some frictions that policy makers should be aware of when formulating IP rights.

*Adopt a "one size fits all" strategy:* when formulating IP rights, policy makers may consult other countries, sectors and organizations. While this may be a helpful approach, policy makers should be careful not to leverage contents that are inconsistent with national and institutional interests and requirements. Moreover, since each country and sector has its own interests and requirements, the cost and benefits of the attack should be different accordingly. Therefore, using a "one size fits all" strategy can be not as effective as well-tailored strategies for the actors. Although policy makers can mitigate this risk by prioritising perceived threat if there is desire to have consistency with the national, international and institutional cybersecurity strategies.

*Neglect links with other national or international strategies:* As we argued in the introduction, IP theft attacks are considered as cyber espionage operations. National [24, 30] and international cybersecurity strategies provide details on how these threats and challenges will be managed. Establishing links across these strategies and sector-specific strategies is not straightforward and this inconsistency makes the identification of essential resources to achieve the strategic objectives listed at national and institutional levels much difficult.

*Lack of an update/review mechanism:* adoption of digital patient records, automation of clinical systems, ease of distributing protected information, heterogeneous nature of networked systems and antiquates clinical applications that are not designed to securely operate in today's networked environment are all security threats in the U.S. healthcare. Evolving threat landscape is also at the core of this increased cybersecurity risk to this sector. According KPMG survey, institutions in this sector may be overconfident about their cybersecurity capabilities [4]. It is probable that the current regulations and agreements could not be responsive to these rapid changes and become irrelevant with the passage of time. Therefore, it is required to have mechanisms to review and update the regulations, agreements and policies among the actors at all levels.

*Lack of an inter-level coordination group:* Since we are living in an interconnected, interdependent digital world, the formulation of IP rights requires input from a variety of governmental, organizational and individual actors. This input can be obtained through different ways. In support of this process, establishment of an inter-level coordination group helps harmonize varying requirements across the actors. This group can also help to translate the technical requirements at the working level into policy-related decisions.

*Failing to identify critical assets:* The protection of critical infrastructures and industries is a common requirement identified by the U.S. National Cybersecurity Structure [30]. There are different studies to identify these critical infrastructures and industries. Similar studies should also be done to identify the most critical type of intellectual properties for specific nations and industries. Prioritizing these IPs is beneficial in planning rapid response in the event of stealing and also in optimizing the costs of the protections.

*Lack of awareness – especially among policy-makers:* a well-developed IP right should provide the actors at the other levels with guidance of concerning key goals, required resources, and how these could be employed most effectively. In the case of a IP right covering a specific industry, raising awareness among the policy makers may be significantly important to facilitate implementation and execution of them.

**Strategic Level**. The advent of cyberspace has changed the business models of the organizations and the interaction among them. These changes may give rise to tensions among the actors depending on the context and preconditions. We know that the context of intellectual property protection is not restricted to a specific entity. The strategic objectives, therefore, must reflect the underlying basics of the processes that provide actual means and identifies the actors involved. Here, we discuss a number of lessons that we learnt from other strategy development models like developing conflict prevention frameworks.

*From complexity to perplexity:* the need to work within a complex environment defined by increased diffusion of power is probably the greatest challenge to the actors. An abundance of actors that need to be engaged in IP protection represents a conceptual and resource challenge to the governments and industries. On one hand, this engagement builds trust among the actors that only a short list of them are collected in Table **??**. On the other hand, the complexity of this environment is not comprehensible for the all actors and may cause confusion and perplexity which unable them to deal with the challenges.

*Unclear communication:* specific definitions, legal frameworks, and implications are inevitable challenges among the industries and nations, let alone the international actors. It is necessary to be aware of this challenge and not let a actor be constrained by particular language of the group in which one is operating. These actors are in danger of ignorance if their communication approach is not aligned with the group.

**Operational Level**. As the link between strategic objectives and tactical employment of forces, activities at operational level of a digital ecosystem can be further expanded due to the importance of coordinating large unit actions and the necessity of synchronization close and deep operations. In the following, we will outline five cases that could have a detrimental effect on effectiveness of a digital ecosystem.

*Leaving a policy vacuum:* policy vacuums occur, intentionally or unintentionally, when there is an absence of specific policies relating to a specific situation at operational level. Often, either no policies for conduct in these situations exist or existing policies seem inadequate. These vacuums will progressively fill itself due to function creep both vertically and horizontally, leading to friction with other public and private organisations, and could also lack proper accountability. A Computer Emergency Response Team (CERT), for example, shall be focused on incident response and recovery. However, when this operational entity lacks a proper strategic vision, function creep may occur towards proactive and prevention aspects of defense.

*Uncoordinated operations:* Cyberspace has become a fifth dimension of warfare, in addition to land, sea, air, and space; countries should therefore prepare accordingly, by establishing standard procedures for defending against, and recovering from, related attacks. Countries are vulnerable to attacks that target strategic assets and infrastructure, and disrupt internet traffic, as a means to create unrest. These threats are exacerbated by a lack of coordination among the actors at all levels. Conflicting laws and regulations also confuse organizations and increases the risks of cyber attacks.

*Drafting obsolete legislation:* As cyber threats evolve, public and private organizations must stay up-to-date. Lawmakers are struggling to keep up with the realities of 3D-printed weapons, for example, and drone interference with aircraft, not to mention the staggering rate of change in terms of increased global internet use and hyper-connectivity of devices. An inability to stay ahead of the curve creates the risk of abuses (through ignorance or malice) with no legal recourse - and may leave entire nations without the ability to make an informed response to cyber attacks.

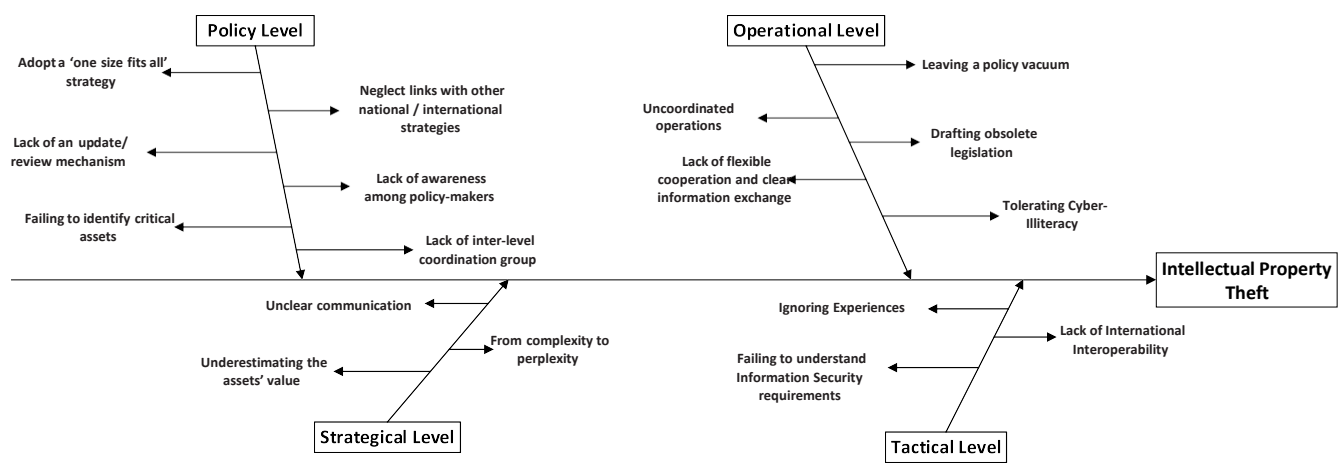

**Figure 1: Casual factors of APT41 in the form of fishbone diagram**

*Lack of flexible cooperation and clear information exchange:* Global threats necessitate new ways of working together to bolster data security. By better connecting academic researchers, private sector organizations, non-profits and government agencies, the transmission and shared analysis of big data can lead to more fruitful discoveries. However, data collaboration will only remain viable if it can be done securely. This entails data anonymization that minimizes so-called re-identification, and international collaboration among lawmakers and law enforcement agencies. As the cyber world increasingly inserts itself into to the physical world, there is a hunger for new ways to collaborate.

*Tolerating cyber-illiteracy:* the existing gap in understanding of cybersecurity issues by higher level officials, decision makers, and policy makers leads to neglect of serious threats and incidents. The unique characteristics of organized cyber crimes such as technology and skill intensiveness, higher degree of globalization and newness urges to close this gap not only among the practitioners, but also among the judges to prevent imbalanced sentences [18].

**Tactical Level.** Good tactics, well-translated strategic objectives into tactical plans to achieve them, enable us to find advantages on adversaries to defeat their malicious operations. With clear understanding of the tactical level activities and avoiding following bad practices, these activities might yield a desired strategic end.

*Ignoring experiences:* A US-China trade deal on IP would not be unprecedented. The two countries concluded a bilateral agreement on IP in 1992, after the United States threatened to increase tariffs on Chinese products [22]. Also, in 2015, President Obama also made a cybersecurity agreement with China to refrain from cyber-enabled theft of intellectual property for commercial gain [9]. Usually, the initial legal and organizational responses to a specific cyber crime will turn out to be inadequate, or obsolete in the face of technical and social changes. Identifying underlying issues, sharing experiences and enhancing national and international interoperability can help to combat evolving and ever-changing cyber threats.

*Failing to understand information security requirements:* Providing cybersecurity is impossible without implication of adequate Information Security Management Systems across the digital ecosystem. For some businesses and sectors (e.g., healthcare) this can be one of the most difficult tasks to accomplish and considered as a disruption due to the traditional practices. However, without meeting the requirements, it is impossible to provide even the most basic protection of the digital systems.

Figure 1 illustrates all the mentioned causal factors that contribute in IP theft in the form of a fishbone diagram. Fishbone diagram, as one of the primary techniques used to perform a five whys analysis, is a basic tool to show causes of a specific event proposed by Kaoru Ishikawa [15, 25].

## Find the root causes

We use the Security by Consensus (SBC) model as a holistic approach considering both social and technical contexts to examine the root causes of this case. The SBC model was first developed in 1989 as a conceptual framework to analyze and compare European and North American technical computer security evaluation criteria [17]. As illustrated in Figure 2, SBC model proposes two general classes of barriers to crime activities: social and technical. This model assumes a potential perpetrator, a suitable victim, and inadequate protection of the victim. In our case, perpetrator and victim are APT41 Chinese state-sponsored group and the U.S. cancer research centers, respectively. Inadequate protection can be classified according to the two classes of barriers or protection mechanisms: inadequate social protection or inadequate technical protection. The opportunity structure for this case is determined by the interplay between criminals, victims and these barriers.

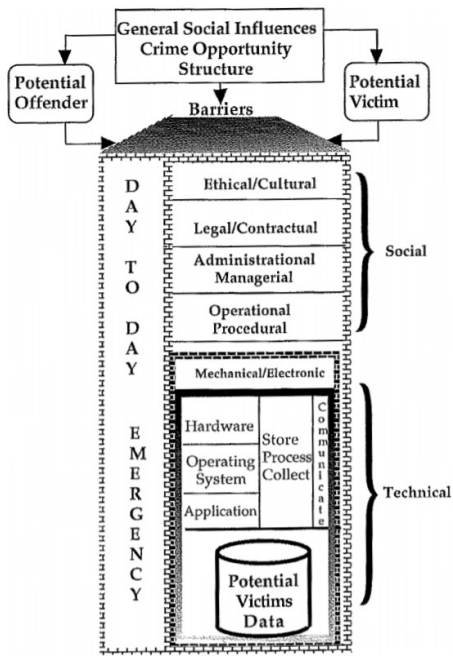

**Figure 2: SBC framework for IT crime prevention**

In the SBC framework, socio-technical systems' actors create circumstances and potentials to become a possible victim of a cyber crime by inadequately protecting the digital systems they own or use. The two general classes of protection measures are social and technical which can be subdivided into other categories. Description of the different categories are given below. The original intention of SBC model is to serve as an abstract model of systems' problems at different levels. As was saw in Figure 2, the SBC framework focuses on both offender and victim's social and technical barriers. This is the strength of this framework to investigate the root causes of this case.

*Cultural/Ethical.* This category encompasses the social behavior and norms found in societies and organizations, as well as the knowledge, values, capabilities and habits of the

individuals in these groups. From cultural and ethical perspective, identified root causes include:

- The U.S.:
  - A shortage of awareness and poor cybersecurity culture in healthcare sector
  - Intellectual property illiteracy in healthcare sector to handle IP rights properly
  - Lack of a counterintelligence mindset, and adversarial and system thinking among the practitioners and decision makers
- China:
  - Adoption of a top-down innovation model (and mindsets)
  - Inadequate education with regard to IP rights

*Legal/Contractual.* Social and governmental institutions use a system of rules to regulate behavior. These legal systems regulate and ensure that individuals or organizations adhere to the will of the state and comply with the regulations. In the context of cybersecurity and intellectual property rights, understanding the unique meaning of types of compliance is crucial in managing risks in an organization:

- Statutory cybersecurity and privacy requirements are required by law and refer to current laws that were passed be a state of federal government.
- Regulatory cybersecurity and privacy requirements are required by law but are different from statutory requirements in that these requirements refer to rules issued by a relating body that is appointed by a state or federal government. These are legal requirements through proxy, where the regulating body is the source of the requirement. It is important to keep in mind that regulatory requirements tend to change more often than statutory requirements.
- Contractual cybersecurity and privacy requirements are required by legal contract between private parties. This may be as simple as a cybersecurity or privacy addendum in a vendor contract that calls out unique requirements. It also includes broader requirements from an industry association that membership brings certain obligations

From legal and contractual perspective, identified root causes include:

- The U.S.:
  - Expensive attorneys fees
  - Lack of perception (among law practitioners) of value that technology is creating
  - Lack of understanding and transparency of the aforementioned compliance
  - Lack of synchronizations of laws with the speed of technology advance

- China:
  - Lack of meaningful injunctive relief
  - Low damage awards
  - Interpretation of IP laws can be fragmented and out of sync with international standards
  - Not being a member of Patent Law Treaty

*Managerial/Administrative.* Management is the administration of an organization at each level of a socio-technical system, and includes the activities of setting the strategy of the organization and coordinating the efforts of its internal actors (e.g. employees, managers, owners, etc.) and external actors (e.g. customers, government, society, suppliers, etc.) to accomplish the objectives through the application of available resources. From managerial and administrative perspective, identified root causes include:

- The U.S.:
  - Incapability of confronting China's incentive plans
  - Insufficient administrative enforcement
  - Insufficient systematic and strategic approach to deploy IP rights and patentability requirements
  - Insufficient tax incentives for the creation of IP assets
- China:
  - Encouraging and assisting in the conduct of cyber attacks
  - Low quality of innovation
  - Inefficient capital markets and resource allocation
  - Slow transformation of state-owned enterprises
  - Difficult and inefficient technology transfer, market access, licensing, and the effective commercialization of IP remain in place

*Operational/Procedural.* Operations and procedures describe the sequence of steps, and specify for each steps what tasks need to be executed and by whom. These steps are aimed at harvesting value from assets owned by the organization. From operational and procedural perspective, identified root causes include:

- The U.S.:
  - Non-established procedures for asset valuation (cost-benefit analysis)
  - Non-established risk assessment procedures
  - Insufficient communication and information exchange procedures inter- and intra-organizations
  - Complacency with cybersecurity practices and lack of effective and timely disabling of access to websites and services whose primary function is to offer infringing content online
  - Failure to conduct third party due diligence
- China:
  - Insufficient action, with low transparency, against the online piracy ecosystem

*Technical.* Technology is the sum of techniques, tools, and processes used in the production of physical devices, services, software platforms and applications to accomplish the organizational objectives. Technology can be embedded in machines to allow for operation, or it can be the knowledge of techniques. From technical perspective, identified root causes include:

- The U.S.:
  - Lack of phishing tests in certain institutions
  - Pervasiveness of legacy systems
  - Use of company-sanctioned services and devices
  - Lack of using Data Loss Prevention (DLP) tools
- China:
  - Growing number of platforms that facilitate online piracy

## Find the solutions

According to a report by Netwrix [23], malware, human error and privileged abuse are the leading intellectual property theft scenarios. Analysis of this case also showed that these were the main factor of attacker's success. This finding shows that to protect IP, or any valuable asset, against attacks, healthcare industries need to mount both social and technical defenses. In this section, discuss three types of corrective actions:

- Immediate action to quickly mitigate the risk of the incident;
- Permanent action to eliminate the root causes; and,
- Preventive action to prevent the incident from recurring.

While permanent actions confirm that the selected correction actually targets the root causes, preventive actions implement any changes to systems, processes, or procedures necessary to prevent recurrence the problems. These actions can be taken at different steps in cybersecurity value chain. In this study, we used the proposed cybersecurity value chain proposed by NIST [3] and suggested 62 corrective actions to tackle this case. To measure the feasibility and effectiveness of these actions, we employed Q Methodology and used convenience sampling accorded five experts (two lawyers, one economists, and two IT professionals) viewpoints. The difference between Q methodology and survey, interview, or focus group approaches is that the response variable in Q methodology is the participant in the study, not the participants' answers to a series of questions [20]. Table 3 indicates top 30 of most feasible and effective actions and their corresponding strategies as selected by the five experts. These results show the importance of Response and Recovery actions for the case APT41 as they are mostly categorized in Immediate corrective actions.

**Table 3: Proposed corrective actions for this case. I: Immediate, Pe: Permanent, Pr: Preventive**

| Strategies | Operations | Tactics |
|---|---|---|
| Deter | Asset Management | Prioritizing assets based on their classification, criticality and business value **(I)** |
| | | Creating RACI matrix for the entire ecosystem **(I)** |
| | Business Environment | Prioritizing organizational objectives and activities **(Pr)** |
| | Governance | Setting up and communicating actor's cybersecurity policy **(Pr)** |
| | Risk Assessment | Identifying and documenting assets' vulnerabilities, internal and external threats, and potential business impacts and likelihood **(Pe)** |
| | | Identifying and prioritizing risk responses **(I)** |
| | Risk Management | Making sure that risk management processes are established and agreed by all actors **(Pe)** |
| | | Determining organizational risk tolerance **(I)** |
| | Supply Chain Risk Management | Making sure that supply chain risk management is established, agreed and conducted by all actors **(Pr)** |
| | | Making sure that response and recovery planning are conducted with suppliers and third parties **(Pr)** |
| Protect | Awareness and Training | Informing all users of their roles and responsibilities and training them **(Pe)** |
| | Data Security | Making sure that data-in-rest and data-in-transit are protected **(I)** |
| | Information Protection Processes | Creating a baseline configuration of IT/IC systems incorporating security principles **(Pr)** |
| | | Creating and testing response plans (Incident Response and Business Continuity) and recovery plans (Incident Recovery and Disaster Recovery) **(Pe)** |
| | Maintenance | Approving and performing physical and online maintenance of the assets **(Pe)** |
| Detect | Anomalies and Events | Analyzing detected incidents to understand attack targets and methods **(Pr)** |
| | | Determination of incidents impacts and alert thresholds **(I)** |
| | Detection Processes | Testing detection processes and making sure they comply with the requirements and continuously improving **(Pr)** |
| Response | Response Planning | Executing response plan during or after an incident **(I)** |
| | Communications | Make sure the personnel understand their roles and order of operations when a response is needed **(Pe)** |
| | | Reporting and sharing incidents consistent with the criteria, response plan and requirements **(I)** |
| | | Coordination with actors according to response plan **(I)** |
| | Analysis | Investigating and categorizing the notifications from detection systems **(I)** |
| | | Performing forensics **(I)** |
| | | Establishing processes to receive, analyze and respond to vulnerabilities disclosed to the organization from internal or external sources **(I)** |
| | Mitigation | Containing and mitigating incidents **(I)** |
| | | Mitigating newly identified vulnerabilities or documenting them as accepted risks **(I)** |
| Recovery | Recovery planning | Executing recovery plan during or after a cybersecurity incident **(I)** |
| | Communications | Managing public relations **(I)** |
| | | Communicating recovery activities with other actors **(I)** |

Operational exercises, exchanging of lessons learned and emphasis on continuous update and review of policies and strategies play significant roles in prevention of cyber-enabled IP theft. This is not limited to technical practitioners and covers a broad range of people who are involving in the provision of cybersecurity. For both professional and non-professionals, cyber ranges can provide a secure environment for cybersecurity education, training and testing [16]. Cyber ranges can build a foundation for such environments relying on both technology and human intelligence. This combination enables us to gain insight into cybercriminals' behavior and raise awareness among the defenders. With proper training provided by the cyber ranges, actors can use this intelligence to identify goals and the underlying processes to achieve them. By that means, better decisions will be made and it is possible to significantly reduce the success rates of the attacks.

## 4 CONCLUSION

Intellectual property is an enabler of innovation and economic growth. Protection of IP rights enables countries and IP-intensive industries to foster development and invest in RD activities. The U.S. is incurring billions of dollars annually due to IP theft. In recent years, in continue of a long-standing approach, countries carry out espionage operations against other countries. In August 2019, Chinese state-sponsored hackers were revealed to have targeted multiple U.S. cancer institutes to take information relating to cutting edge cancer research. This is one of the industries targeted by APT41, a prolific Chinese cyber threat group that carries out state-sponsored espionage activity in parallel with financially motivated operations. Reports show that this group started their activities from 2013. In this paper, we conducted a socio-technical root cause analysis to examine the casual factors and root causes of these operations.

The aim of a root cause analysis is to find the main reasons of emerging a problem and proposing actions to eliminate these problems. Our study shows that the economic benefit of preventing IP theft in the U.S. is significant. On the other side, because of large productivity gains in China, expecting the implementation of mutually beneficial IP rights protection by the Chinese government is not a subtle strategy to adopt by the U.S. Assessing the significance of each political, economic, socio-cultural and technological factors on the prevention, response and recovery from these incidents necessitates new analytical studies. However, this study showed that most of the root causes of APT41 are implanted in the U.S. industries and institutions. These findings complement the results presented in Table 3 that the top immediate and permanent actions selected by the experts focus on the U.S. In this table, we proposed a set of corrective actions that, if followed precisely, prevent recurring this problem. We evaluated the effectiveness and feasibility of these action by conducting a systematic study of experts viewpoints. As a result, this study outlined thirty actions that are suggested to operationalize immediately and permanently. Due to the multidisciplinary nature of this problem, these corrective actions are not limited to technical solutions (e.g. and covers the social solutions as well. Hence, we believe that professionals and non-professionals can leverage the educating and training capabilities that cyber ranges can provide for individuals, organizations and governments and build a resilient infrastructure.

In an effort to better understand the selected case and define the problem, we used the qualitative and quantitative evidence drawn from a wide range of sources such as primary and secondary legislation, reports from government agencies, legal analysis by practitioners, studies from international organizations (e.g. OECD, WTO, and WIPO), academic and legal journals, and news websites. All sources used are publicly available and are free and accessible to all.

## REFERENCES

[1] Christian Aghroum. 2012. Foreign spies stealing US economic secrets in cyberspace. Report to Congress on foreign economic collection and industrial espionage. 2009-2011. *Securite et strategie* 8, 1 (2012), 78–79.

[2] Bjørn Andersen and Tom Fagerhaug. 2006. *Root cause analysis: simplified tools and techniques.* Quality Press.

[3] Matthew P Barrett. 2018. *Framework for improving critical infrastructure cybersecurity version 1.1.* Technical Report.

[4] G Bell and M Ebert. 2015. Health care and cyber security: increasing threats require increased capabilities. *KPMG* (2015).

[5] Global Innovation Policy Center. 2019. Inspiring Tomorrow, US Chamber International IP Index (7th Edition).

[6] IP Commission. 2017. The Report of the Commission on the Theft of American Intellectual Property. http://ipcommission.org/report/IP_Commission_Report_Update_2017.pdf

[7] Carlos M Correa. 2016. The Current System of Trade and Intellectual Property Rights. In *European Yearbook of International Economic Law 2016.* Springer, 175–197.

[8] Kaspersky Daily. 2018. Top 5 most notorious cyber-attacks. https://www.kaspersky.com/blog/five-most-notorious-cyberattacks/24506/

[9] Julie Hirschfeld Davis and David E. Sanger. 2015. Obama and Xi Jinping of China Agree to Steps on Cybertheft. https://www.nytimes.com/2015/09/26/world/asia/xi-jinping-white-house.html

[10] A Mark Doggett. 2005. Root cause analysis: a framework for tool selection. *Quality Management Journal* 12, 4 (2005), 34–45.

[11] International Agency for Research Cancer. 2018. China's Fact Sheet. https://gco.iarc.fr/today/data/factsheets/populations/160-china-fact-sheets.pdf

[12] Chuck Grassley. 2019. Grassley on Chinese Espionage: It's called cheating. And it's only getting worse. https://www.judiciary.senate.gov/grassley-on-chinese-espionage-its-called-cheating_and-its-only-getting-worse

[13] The White House. 2018. Statement from the Press Secretary Regarding the President's Working Dinner with China. https://www.whitehouse.gov/briefings-statements/statement-press-secretary-regarding-presidents-working-dinner-china/

[14] Nancy Hungerford. 2019. Chinese theft of trade secrets on the rise, the US Justice Department warns. https://www.cnbc.com/2019/09/

23/chinese-theft-of-trade-secrets-is-on-the-rise-us-doj-warns.html

[15] Kaoru Ishikawa. 1982. *Guide to quality control.* Number TS156. I3713 1994.

[16] Mazaher Kianpour, Stewart Kowalski, Erjon Zoto, Christopher Frantz, and Harald Øverby. 2019. Designing Serious Games for Cyber Ranges: A Socio-technical Approach. In *2019 IEEE European Symposium on Security and Privacy Workshops (EuroS&PW).* IEEE, 85–93.

[17] Stewart Kowalski. 1996. IT insecurity: A multi-disciplinary inquiry. (1996).

[18] Nir Kshetri. 2010. *The global cybercrime industry: economic, institutional and strategic perspectives.* Springer Science & Business Media.

[19] James Andrew Lewis. 2013. Significant cyber incidents since 2006. *Center for Strategic and International Studies* (2013).

[20] Aaron M Lien, George B Ruyle, and Laura Lopez-Hoffman. 2018. Q Methodology: A method for understanding complex viewpoints in communities served by extension. *Journal of Extension* 56, 2 (2018), 2IAW4.

[21] Keith E Maskus. 1998. The role of intellectual property rights in encouraging foreign direct investment and technology transfer. *Duke J. Comp. & Int'l L.* 9 (1998), 109.

[22] Jean-Frédéric Morin and Dimitri Thériault. 2019. Copyright Provisions in Trade Deals: A Bird's-eye View. (2019).

[23] Netwrix. 2019. 2018 IT Risks Report. http://www.netwrix.com/go/research

[24] Cyberspace Administration of China. 2016. Full text of National Cyberspace Security Strategy. cac.gov.cn/2016-12/27/c_1120195926.htm

[25] Taiichi Ohno. 1988. *Toyota production system: beyond large-scale production.* crc Press.

[26] World Health Organization Representative Organization. 2019. The situation in China. http://www.wpro.who.int/china/mediacentre/factsheets/cancer/en/

[27] South China Morning Post. 2019. FBI has 1,000 investigations into Chinese intellectual property theft. https://www.scmp.com/news/china/article/3019829/fbi-has-1000-probes-chinese-intellectual-property-theft-director

[28] Samantha F Ravich. 2015. *Cyber-Enabled Economic Warfare: An Evolving Challenge.* Hudson Institute.

[29] Eric Rosenbaum. 2018. 1 in 5 corporations say China has stolen their IP within the last year: CNBC CFO survey. https://www.cnbc.com/2019/02/28/1-in-5-companies-say-china-stole-their-ip-within-the-last-year-cnbc.html

[30] Donald Trump. 2018. National Cyber Strategy of the United States of America. *Washington, DC* (2018).

[31] GC Wilshusen. 2012. *Information Security: Cyber Threats Facilitate Ability to Commit Economic Espionage., Pub. L. No.* Technical Report. GAO-12-876T.